\def\doublespace{\baselineskip=\normalbaselineskip 
\multiply\baselineskip by 2}

\doublespace

\def\frac#1/#2{\leavevmode\kern.1em
\raise.5ex\hbox{\the\scriptfont0 #1}\kern=.1em
/\kern-.15em\lower.25ex\hbox{\the\scriptfont0 #2}}

\documentstyle[preprint,aps]{revtex}

\begin{document}
\draft
\preprint{\today}
\title
{Quasiparticle Lifetimes in Superconductors: Relationship with the
Conductivity Scattering Rate}
\def \bk{{\bf k}}
\def \bQ  {{\bf Q}}
\def \bp {{\bf p}}
\def \bq {{\bf q}}
\def\wtilde{\tilde {\omega }}
\def\YBCO{YBaCu$_3$O$_{7-x}$}
\def\BKBO{Ba$_{1-x}$K$_x$BiO$_3$}
\def\bea {\begin{eqnarray}}
\def\eea {\end{eqnarray}}
\def\be {\begin{equation}}
\def\ee {\end{equation}}

\def\bl {{\bf l}}
\def \Dl {\Delta_l}
\author{F. Marsiglio$^{1,2,3}$ and J.P. Carbotte$^{2,3}$}
\address
{$^1$Neutron \& Condensed Matter Science\\
AECL, Chalk River Laboratories, Chalk River, Ontario, Canada K0J 1J0\\
$^2$Dept. of Physics \& Astronomy, McMaster University, 
Hamilton, Ontario L8S 4M1 \\
$^3 $ Canadian Institute for Advanced Research, McMaster University,
Hamilton, ON L8S 4M1}

\maketitle
\begin{abstract}
We compute the single particle inverse lifetime, evaluated in the
superconducting state. Within the BCS framework, the calculation can
be done non-perturbatively, i.e. poles can be found well away from the
real axis. We find that perturbative calculations are in good agreement
with these results, even for non-zero impurity scattering. 
With
electron-phonon scattering added to the problem, we use the Eliashberg
equations, with impurity scattering, to calculate the quasiparticle
inverse lifetime perturbatively. In all cases we find that the 
inverse lifetime
is significantly suppressed in the  superconducting state,
particularly in the presence of impurity scattering. We also compute the 
far-infrared and microwave conductivity, and describe procedures for extracting 
scattering rates from conductivity measurements. While these procedures
lead to scattering rates in qualitative agreement with the inverse lifetime,
we find that quantitative agreement is lacking, in general.

\end{abstract}
\vfil\eject
\bigskip
\noindent {\bf I.  INTRODUCTION}
\par
The quasiparticle lifetime is a concept which is useful 
for clarifying the nature of 
the system of interest, i.e. is it a Fermi vs. marginal Fermi vs. a 
Luttinger liquid, etc \cite{1}. Various techniques are available for measuring
lifetimes \cite{kaplan}; perhaps the most direct is through tunnel
junction detection \cite{narayanamuti}. In this paper we analyze
electron scattering rates,  as measured by microwave
and far-infrared conductivity measurements, and examine their relationship to
quasiparticle lifetimes.
The experimental results for
\YBCO \cite{bonn1,bonn2} and their theoretical implications
\cite{bonn2,berlinsky,klein,bonn3} have been discussed extensively in 
the literature. We include a treatment of inelastic scattering
(through the electron-phonon interaction)  and use Eliashberg theory
to further analyze the problem.
To some extent this has been done in the normal state already
\cite{shulga,marsiglio95a}. 
Certain aspects of quasiparticle lifetimes are clarified
and the analysis is extended to the superconducting state,
following our recent work on the
optical conductivity \cite{marsiglio95a,marsiglio95b}.\par

We will proceed by defining what we mean by a single particle lifetime,
bearing in mind that sometimes such a definition is not very meaningful,
or useful. We review two ways in which conductivity data can
be used to extract a quasiparticle scattering rate, one based on the two-fluid
model, and the other based on a straightforward Drude fit to the low 
frequency conductivity. The instances in which these procedures
yield a qualitative or quantitative facsimile of the quasiparticle inverse
lifetime will be clarified. In this way, one can evaluate the usefulness of the
two-fluid hypothesis, for example, in systems with both elastic and inelastic
scattering channels.\par

This paper has been divided into
two sections. The first reviews quasiparticle lifetimes, which are 
calculated from the single particle Green function. Previous work is extended
to include both electron-impurity scattering (in the Born approximation)
and electron-phonon scattering. Perturbative methods within the Eliashberg 
framework are used. The second section elucidates two methods recently
used to extract scattering rates from optical and microwave conductivity
measurements, and how these relate to the single particle inverse lifetime.
We should make it clear at the start that the calculation we use for
the optical conductivity omits vertex corrections. The main result
of this is that the ``$1 - \cos{\theta}$'' factor that occurs in a
Boltzmann formulation of transport properties is absent \cite{mahan},
so in fact, when theoretical comparisons between the quasiparticle
lifetime and the conductivity-derived scattering rate are made, the
agreement will in general be better than it would be, given an exact
calculation. This is made obvious in the next section, which treats
elastic impurity scattering. Conversely, good agreement between these
two quantities in the theory does not imply good agreement between the
two measured quantities.

\par
\noindent {\bf II.  QUASIPARTICLE LIFETIMES}
\par
\medskip
\noindent {\bf II.A  Normal State}
\par
\medskip
The fact that the quasiparticle lifetime is related to the conductivity is
most apparent in
the Drude model, which incorporates electron-impurity scattering into the
problem. In the Born approximation, the single-electron self-energy is given
by
\begin{equation}
\Sigma(\omega+i\delta) = -{i \over 2 \tau}
\label{impurityself}
\end{equation}
\noindent where $1/\tau$ is the impurity scattering rate and $\tau$ is
the single particle lifetime. We will utilize
an expression for the optical conductivity which omits vertex corrections,
and so it is given simply as the convolution of two single particle Green
functions \cite{nam,lee,bickers,marsiglio92}. The result reduces to
the Drude conductivity \cite{ashcroft},
\begin{equation}
\sigma(\nu) = {ne^2 \over m} {1 \over 1/\tau - i \nu},
\label{drude}
\end{equation}
\noindent whose real part is given by the usual Lorentzian form with
half-width $1/\tau$. 
Note that vertex corrections would actually alter
the $1/\tau$ that appears in eq. (\ref{drude}); however, if only s-wave
phase shifts are included, the two inverse lifetimes would be
identical.
Here $n$, $e$, and $m$ are the free electron density,
charge magnitude, and mass. Thus the single particle 
lifetime appears explicitly
(and can easily be measured) in the optical conductivity.\par

The next step is to include phonons, and thus an additional scattering 
mechanism, via the electron-phonon interaction \cite{allen,shulga}.
Then the electron self-energy is given by \cite{allenmitrovic}
\begin{equation}
\Sigma(\omega + i \delta) = \int_0^\infty d\nu \alpha^2F(\nu)
\Biggl\{ -2\pi i \bigl[N(\nu) + {1 \over 2} \bigr] + \psi({1\over 2}
+i{\nu - \omega \over 2\pi T}) - \psi({1\over 2}
-i{\nu + \omega \over 2\pi T})
\Biggr\} - {i \over 2\tau}, 
\label{phononself}
\end{equation}
\noindent where $\alpha^2F(\nu)$ is the electron-phonon spectral function,
$N(\nu)$ is the bose function, and $\psi(x)$ is the digamma function. The 
quasiparticle has energy and lifetime defined by the pole of the single 
particle retarded Green function, i.e. the zero of
\begin{equation}
G^{-1}(k,\omega + i\delta) = \omega - \epsilon_k - \Sigma(\omega + i \delta),
\label{zero}
\end{equation}
\noindent where
now the functions are analytically continued to the lower half-plane.
Defining the energy and inverse lifetime by
\begin{equation}
\omega \equiv E_k - i \Gamma_k
\label{pole}
\end{equation}
\noindent we obtain
the two equations,
\begin{equation}
E_k = \epsilon_k + \int_0^\infty d\nu \alpha^2F(\nu)
Re \Biggl\{
\psi({1\over 2} + i{\nu - E_k + i\Gamma_k \over 2\pi T}) - 
 \psi({1\over 2} - i{\nu + E_k - i\Gamma_k \over 2\pi T})
\Biggr\}
\label{real}
\end{equation}
\begin{equation}
\Gamma_k = {1\over 2\tau} + \int_0^\infty d\nu \alpha^2F(\nu)
\Biggl\{
2\pi \bigl[N(\nu) + {1 \over 2} \bigr] - Im \Bigl\{
\psi({1\over 2} + i{\nu - E_k + i\Gamma_k \over 2\pi T}) - 
\psi({1\over 2} - i{\nu + E_k - i\Gamma_k \over 2\pi T}) \Bigr\}
\Biggr\}.
\label{imag}
\end{equation}
\noindent The general problem requires a numerical solution of these equations
and is beyond the scope of this paper.
Here, we assume that $E_k$ and $\Gamma_k$ are small so that
one can obtain closed expressions. They are \cite{allenmitrovic,grimvall}
\begin{equation}
E_{k} = { \epsilon_k \over 
1 + \lambda^\ast(T) }
\label{enerk}
\end{equation}
\begin{equation}
\Gamma_{k} = { {1\over 2\tau} + 2\pi\int_0^\infty d\nu \alpha^2F(\nu)
\bigl[N(\nu) + f(\nu) \bigr] \over 1 + \lambda^\ast(T) }
\label{gammak}
\end{equation}
\noindent where
$f(\nu)$ is the Fermi function and
\begin{equation}
\lambda^\ast(T) = -{1\over \pi T} \int_0^\infty d\nu \alpha^2F(\nu)
Im \psi^\prime({1 \over 2} + i{\nu \over 2 \pi T}).
\label{lambdastar}
\end{equation}
As $T\rightarrow 0$, $\lambda^\ast(T) \rightarrow \lambda \equiv
2\int_0^\infty {d\nu \over \nu}\alpha^2F(\nu)$, while at high temperature,
$\lambda^\ast(T) \propto 1/T^2$. Typical results for various electron-phonon
spectral functions are shown in Fig. 1.

\medskip
\noindent {\bf II.B  Superconducting State}
\par
\medskip
Quasiparticle lifetimes in electron-phonon superconductors were discussed
long ago by Kaplan {\it et al.} \cite{kaplan}. Before proceeding to their
relationship with the optical conductivity we give a brief review of the
formalism \cite{kaplan,nicol}. As in the normal state the quasiparticle 
scattering rate is defined by (twice) the imaginary part of the pole in
the single particle Green function. In the superconducting state, the
diagonal component of the single particle Green function is \cite{scalapino69}
\begin{equation}
G_{11}(k,\omega) = { \omega Z(\omega) + \epsilon_k \over
\omega^2 Z^2(\omega) - \epsilon_k^2 - \phi^2(\omega) },
\label{green}
\end{equation}
\noindent where
$Z(\omega)$ and $\phi(\omega)$ are the renormalization and pairing functions
given by solutions to the Eliashberg equations \cite{eliashberg,scalapino69,msc}
which are repeated here for convenience:
\begin{eqnarray}
\phi(\omega) & = & \pi T\sum\limits_{m=-\infty}^{\infty}
\bigl[\lambda(\omega-i\omega_m)-\mu^*(\omega_c)\theta(\omega_c-\vert{\omega_m}
\vert)
\bigr]{\phi_m\over\sqrt{\omega^2_m Z^2(i\omega_m)+\phi_m^2}}
\nonumber \\
 & & +i\pi\int_0^{\infty}d\nu\,\alpha^2F(\nu)\Biggl\{\bigl[N(\nu)+f(\nu-
\omega)\bigr]{\phi(\omega-\nu)\over\sqrt{\wtilde^2(\omega-\nu) -
\phi^2(\omega-\nu)}}
\nonumber \\
 & & +\bigl[N(\nu)+f(\nu+\omega)\bigr]{\phi(\omega+\nu)\over\sqrt{
\wtilde^2(\omega+\nu) - \phi^2(\omega+\nu)}}\Biggr\}
\label{phireal}
\end{eqnarray}
\noindent and
\begin{eqnarray}
\wtilde(\omega) & = & \omega + i\pi T\sum\limits_{m=-\infty}^{\infty}
\lambda(\omega-i\omega_m)
{\omega_m Z(i\omega_m)\over\sqrt{\omega^2_m Z^2(i\omega_m)
+\phi_m^2}}
\nonumber \\
 & & + i\pi \int_0^{\infty}d\nu\,\alpha^2F(\nu)\Biggl\{\bigl[N
(\nu)+f(\nu-\omega)\bigr]{\wtilde(\omega-\nu)\over\sqrt{\wtilde^2(\omega-\nu)
-\phi^2(\omega-\nu)}}
\nonumber \\
& & +\bigl[N(\nu)+f(\nu+\omega)\bigr]{\wtilde(\omega+\nu)\over\sqrt{
\wtilde^2(\omega+\nu)-\phi^2(\omega+\nu)}}\Biggr\}\ ,
\label{wtildereal}
\end{eqnarray}
\noindent where $\wtilde(\omega) \equiv \omega Z(\omega)$.
Here, $N(\nu)$ and $f(\nu)$ are the Bose
and Fermi distribution functions, respectively. The electron-phonon spectral 
function is given by $\alpha^2F(\nu)$, its Hilbert transform is
$\lambda(z)$. The Coulomb repulsion parameter is $\mu^\ast(\omega_c)$
with cutoff $\omega_c$. A negative value for this parameter
can be used to model some BCS attraction of unspecified origin.
The renormalization and pairing
functions are first obtained on the imaginary axis at the 
Matsubara frequencies, i.e. $\omega = i\omega_n \equiv
i\pi T(2n-1)$, with $\phi_m \equiv \phi(i\omega_m)$ by setting the complex
variable $\omega$ in these equations to the Matsubara frequencies 
\cite{owen,rainer}.
Then the equations are iterated as written, with $\omega$ set to a frequency
on the real axis. Note that the square roots with complex arguments are
defined to have a positive imaginary part.\par

Actually, these functions are obtained at frequencies just above the real axis,
and then, in principle, the pole is given by the zero of the denominator 
continued to the lower half-plane. However, since the solutions are
readily known only along certain lines in the complex plane (e.g. the imaginary
axis or just above the real axis) we follow previous authors and look
for the pole perturbatively. That is, we write $\omega \equiv E - i\Gamma$,
and linearize the imaginary part so that \cite{scalapino69,kaplan}
\begin{equation}
\Gamma(E) = { EZ_2(E) \over Z_1(E)} - {\phi_1(E) \phi_2(E) \over EZ_1^2(E)}
\label{lifetime}
\end{equation}
\noindent Note
that $E$ is determined by equating the real parts (again linearizing 
in imaginary components), which yields
\begin{equation}
E = \sqrt{{\epsilon_k^2 + \phi_1^2(E) \over Z_1^2(E)}}.
\label{energy}
\end{equation}
At the Fermi surface, $\epsilon_k \equiv 0$ so $E = \phi_1(E)/Z_1(E)
\equiv \Delta_1(E)$ becomes the definition for the lowest energy 
excitation, i.e. the gap in the excitation spectrum. It has become common
practice to consider $E$ as an independent variable, and then to study
$\Gamma(E)$ as a function of $E$. In Fig. 2 we show $\Gamma(E)$ vs. $E$
for various temperatures, using a Debye model spectrum. It is clear
that the scattering rate near the gap edge (shown by the arrow)
decreases very quickly as the temperature decreases. Note
that $\Gamma(E)$ actually becomes negative at intermediate temperatures
near $E \approx 10$ meV. This is real (i.e. not a numerical artifact)
and is simply a property of the function $\Gamma(E)$ given by 
eq. (\ref{lifetime})
through the linearization procedure. The true pole in eq. (\ref{green}) will
always
have a negative imaginary part (i.e. $\Gamma$ is positive).\par

What happens in the BCS limit ? Then $Z_2(E) \rightarrow 0$ and
$\phi_2(E) \rightarrow 0$, i.e. the functions involved in the solution are
pure real, with $Z_1(E) \rightarrow 1$ and $\phi_1(E) \rightarrow \Delta(T)$,
where $\Delta(T)$ is obtained self-consistently from the BCS equation. Clearly
then the quasiparticle energy is $E = \sqrt{\epsilon_k^2 + \Delta^2}$ and
the scattering rate, $\Gamma = 0$. Thus, within the BCS approximation  the
quasiparticle states are infinitely long-lived, as there is no means by 
which a quasiparticle can decay. In the electron-phonon theory, the
phonons provide the means. In this way the BCS approximation is
pathological.\par

Before including the electron-phonon interaction as a scattering mechanism,
one can first introduce impurities to provide elastic electron scattering. Then
\def\sqq{\sqrt{\omega^2 Z^2(\omega) - \phi^2(\omega)}}
\def\sqqbcs{\sqrt{\omega^2  - \Delta^2}}
\begin{eqnarray}
\phi(\omega) & = & \phi_{cl}(\omega) + {i \over 2\tau} {\phi(\omega) \over\sqq}
\nonumber \\
Z(\omega) & = & Z_{cl}(\omega) + {i \over 2\tau} {Z(\omega) \over\sqq}
\label{impeqn}
\end{eqnarray}
\noindent where
the subscript `cl' refers to the clean limit and $1/\tau$ is the (normal)
impurity scattering rate. The gap function is defined
\begin{equation}
\Delta(\omega) = \phi(\omega)/Z(\omega)
\label{gapfunction}
\end{equation}
\noindent and is independent of impurities. In the BCS limit,
\begin{eqnarray}
\phi(\omega) & = & \Delta + {i \over 2\tau} {\Delta \hbox{sgn} \omega 
\over\sqqbcs}
\nonumber \\
Z(\omega) & = & 1 + {i \over 2\tau} {\hbox{sgn} \omega \over\sqqbcs}.
\label{impeqnbcs}
\end{eqnarray}
\noindent As before,
the quasiparticle energy is given by $E = \sqrt{\epsilon_k^2 + \Delta^2}$,
but the scattering rate is now
\begin{equation}
\Gamma = {1 \over 2\tau} {\sqrt{E^2 - \Delta^2} \over | E | }
 = {1 \over 2\tau} {| \epsilon_k | \over \sqrt{\epsilon_k^2 + \Delta^2}}.
\label{gammaimpbcs}
\end{equation}
\noindent Eq. (\ref{gammaimpbcs}) shows that at the Fermi surface
impurities are ineffectual for electron scattering, i.e. this is another
manifestation of Anderson's theorem \cite{anderson}.
This happens abruptly at $T_c$; as soon as a gap develops the scattering 
rate becomes zero. Away from the Fermi surface the scattering rate is reduced
from its value in the normal state. In particular, well away from the Fermi
surface the scattering rate in the superconducting state approaches the normal
state rate. This behaviour is illustrated in Fig. 3.\par

The calculations in Fig. 3 rely on a perturbative search for the pole in
the lower half-plane, as given by eq. (\ref{lifetime}). A more rigorous
calculation reveals poles ``within the gap'', but these do not appear
in the spectral function because of coherence factors. This can be seen by
rewriting eq. (\ref{green}) as
\bea
G_{11}(k,\omega) & = & \phantom{+}{1 \over 2} \Biggl(1 + 
{\omega \over \sqqbcs}\Biggr) {1 \over
\sqqbcs - \epsilon_k + {i \over 2 \tau} \hbox{sgn} \omega }
\nonumber \\
& & - {1 \over 2} \Biggl(1 - {\omega \over \sqqbcs}\Biggr) {1 \over
\sqqbcs + \epsilon_k + {i \over 2 \tau} \hbox{sgn} \omega }
\label{greennew}
\eea
\noindent The coherence factors, $\Bigl(1 \pm {\omega \over \sqqbcs}\Bigr)$,
are such that the spectral function, $ A(k,\omega) \equiv - {1 \over \pi}
Im G(k,\omega)$, is zero for $| \omega | < \Delta$, independent of the impurity
scattering rate, as can be readily verified explicitly from 
eq.~(\ref{greennew}). On the other hand the pole of the Green function
is given by solving for $\omega = E - i\Gamma$ in the two coupled
equations
\begin{equation}
E^2 - \Gamma^2 = \Delta^2 + \epsilon_k^2 - ({1 \over 2\tau})^2
\label{realpart}
\end{equation}
\begin{equation}
\Gamma = { | \epsilon_k | \over | E | } {1 \over 2\tau}
\label{imagpart}
\end{equation}
\noindent For $1/\tau = 0$ the solution is as before, $E = E_k \equiv
\sqrt{\epsilon_k^2 + \Delta^2}$, and $\Gamma = 0$. For finite impurity
scattering, however, the solution depends on ${1 \over 2\tau \Delta}$ at the 
Fermi surface ($\epsilon_k = 0$):
\begin{eqnarray}
E  =  \sqrt{\Delta^2 - ({1 \over 2\tau})^2}, \phantom{aaaaaaaaa}
\Gamma = 0 \phantom{aaaaaaaaa}\hbox{for} \phantom{aaaa}
{1 \over 2\tau \Delta} < 1
\label{ee1} \\
E  =  0,  \phantom{aaaaaaaaaaa}\Gamma = \sqrt{({1 \over 2\tau})^2 - \Delta^2} 
\phantom{aaaa}\hbox{for}\phantom{aaaa}
{1 \over 2\tau \Delta} > 1
\label{ee2}
\end{eqnarray}
\noindent In either case the solution is irrelevant, because the real part lies
within the gap (i.e. the quasiparticle residue is zero). The relevant energies
are $| E | > \Delta$, and then the scattering rate given by Eq. (\ref{imagpart})
agrees with the linearized solution, Eq. (\ref{lifetime}), although the
quantity $\Gamma$ no longer truly represents a quasiparticle inverse lifetime.
Away from the Fermi surface the quasiparticle energy requires the solution of
a quadratic equation and $E$ may lie below or above the gap, depending on 
$\epsilon_k$ and $1/\tau$. Once more solutions with $| E | < \Delta$ are
irrelevant because the residue is zero. In Fig. 4a (b) we show the
real (imaginary) parts of the pole for some typical parameters. As can be
seen, some atypical behaviour can occur as a function of temperature,
for example when $\epsilon_k = 0$. As the temperature is lowered the pole
moves from a point on the negative imaginary axis to the origin 
(near $T/T_c = 0.8$) and then moves along the real axis towards 
$E = \sqrt{\Delta^2 - ({1 \over 2 \tau})^2}$. Nonetheless, comparison
of Fig. 4(b) with Fig. 3 shows that the perturbative is in qualitative agreement
with the non-perturbative solution.\par

Coherence factors have always been thought to play an important role
for two-particle response functions (such as the NMR relaxation rate
or the microwave conductivity). In particular they lead to the
Hebel-Slichter \cite{hebel} singularity in the NMR relaxation rate.
However they are considered not to play a big role in 
single particle functions.
Eq. (\ref{greennew}) shows, however, that the coherence factors play
a very important role in the single particle spectral function, in
that they maintain a gap equal to $\Delta$, even when impurities
cause the single particle pole to have a real part whose value
lies in the gap.\par

The frequency dependence of the
spectral function is shown in Fig. 5. For energies close to the Fermi surface 
the spectral function is dominated by the square root singularity at the gap 
edge, which comes from the coherence factors rather than the single particle 
pole. For larger energies the peak present is due to the pole and differs
from the normal state spectral function only at low frequencies.\par

We return now to the strong coupling case. In the superconducting state,
we do not have access to the analytic continuation of the Green function
to the lower half-plane \cite{comm}. We thus follow Karakozov {\it et al.}
\cite{karakozov}, and use an expansion near $\omega = 0$. We also follow
Kaplan {\it et al.} \cite{kaplan} and henceforth use Eq. (\ref{lifetime})
for the scattering rate, noting that while the energy $E$ at which $\Gamma(E)$
is evaluated ought, in principle, to be computed self-consistently as was 
done in BCS, i.e. eqs. (\ref{realpart},\ref{imagpart}), in practice 
we will choose the relevant energy, and so treat energy as an independent 
variable.\par

Karakozov {\it et al.} \cite{karakozov} pointed out that at any finite 
temperature the solutions to the Eliashberg equations have the following 
low frequency behaviour:
\begin{eqnarray}
Z(\omega) \approx  Z_1 + {i\gamma_2 \over \omega}
\label{zlow} \\
\Delta(\omega)  \approx  \delta_1 \omega^2 - i \delta_2 \omega
\label{gaplow}
\end{eqnarray}

\noindent where $Z_1, \gamma_2, \delta_1$ and $\delta_2$ are real
(positive) constants. This implies that a quasiparticle pole exists with
\begin{equation}
E \approx {\epsilon_k \over Z_1 \sqrt{1 + \delta_2^2}}
\label{enerlow}
\end{equation}

\begin{equation}
\Gamma \approx {1 \over Z_1} \Bigl( 2\pi \int_0^\infty d\nu \alpha^2F(\nu)
g(\nu) \bigl[ N(\nu) + f(\nu) \bigr] + {g(0) \over 2\tau } \Bigr)
\label{gammalow}
\end{equation}
\noindent where we have used Eq. (\ref{lifetime}) and assumed $E \rightarrow 0$
(i.e. we are near the Fermi surface). In Eq. (\ref{gammalow}) $g(\nu)$ is the
single electron density of states in the superconducting state,
\def\sqqbcsn{\sqrt{\nu^2  - \Delta^2}}
\begin{equation}
g(\nu) = Re \Biggl( {\nu \over \sqqbcsn} \Biggr),
\label{dos}
\end{equation}
\noindent which is non-zero for $\nu = 0$ at any finite temperature
\cite{karakozov,marsiglio91,allenrainer}. In Fig. 6 we show the quasiparticle
scattering rate, $2\Gamma(T,E=0)$ vs. $T/T_c$ for various impurity
scattering rates, $1/\tau$, in the superconducting state. The normal 
state result is also shown for  reference. Note that in the clean limit
there is an enhancement just below $T_c$ in the superconducting state, but
at low temperatures there is an exponential suppression, compared to the
power law behaviour observed in the normal state. When impurity scattering 
is present there is an immediate suppression below $T_c$  (a knee is still
present, however, in the superconducting state). It is clear that at low 
temperatures the superconducting state is impervious to impurity scattering,
as one would expect. \par

Is $2\Gamma(T,E=0)$ the relevant inverse lifetime by which properties
of the superconducting state can be understood ? The answer is no, at 
least not at low temperatures, where the spectral function essentially
develops a gap for low energies. In Fig. 7 we plot the spectral function at
the Fermi surface, $A(k_F,\omega)$ for various temperatures \cite{marsiglio91}.
While no true gap exists at any finite temperature, it is clear that for low
temperatures the spectral weight at low frequency is exponentially suppressed.
Thus we have a situation similar to that in BCS theory, where
the imaginary part of the pole is irrelevant; rather it is the 
scattering rate (as defined by Eq. (\ref{lifetime})) evaluated
above the ``gap-edge'' where a large spectral weight is present, that is
most relevant. When impurities are added the spectral peak is broadened,
even at low temperatures, as shown in Fig. 8 for an intermediate
temperature, $T/T_c = 0.5$. However the lineshape is very asymmetric as a gap
remains at low frequencies (particularly prominent at low temperatures).
Fig. 7 demonstrates that the relevant energy is a function of temperature.
In Fig. 9 we plot $2\Gamma(T,E=\Delta(T))$ vs. $T/T_c$, where $\Delta(T)$
is determined from the relation (at the Fermi surface)
\begin{equation}
E = Re \Delta(E,T).
\label{gap}
\end{equation}
\noindent By Eq. (\ref{gap}) we understand that the $E=0$ solution is excluded
(similarly the very low energy solution is also excluded). We are
interested in the conventional solution which gives rise to the peak in
the spectral function illustrated in Fig. 7 or 8. If no non-zero 
solution is present
(as is the case near $T_c$) then we utilize the $E=0$ solution. Note that
in this case care must be taken when obtaining the $E \rightarrow 0$ limit
of Eq. (\ref{lifetime}). Also there is present in this definition a
discontinuity at some temperature near $T_c$, which is where a nonzero
solution first appears. In this way
we hope to show the scattering rate at an energy where the spectral weight
is large, and therefore of most relevance to observables. At any rate, 
it is clear by 
comparing Fig. 9 to Fig. 6 (see also Fig. 2) that there is very little 
difference in the scattering rate in the gap region of energy. However, as the
energy increases beyond the gap the scattering rate increases, resulting
in short ``lifetimes'', even at zero temperature \cite{kaplan}.
\par
\bigskip
\noindent {\bf III.  THE OPTICAL CONDUCTIVITY}
\par
\bigskip
Using the Kubo formalism \cite{mahan}, the optical conductivity can be
related to a current-current correlation function. The final result for 
the frequency dependence of the conductivity in the long wavelength limit
is  \cite{nam,lee,bickers,marsiglio92}
\begin{equation}
\sigma(\nu) = {i \over \nu} \Bigl( \Pi(\nu + i\delta) + {ne^2 \over m} \Bigr),
\label{cond}
\end{equation}
\noindent where the paramagnetic response function, $\Pi(\nu + i\delta)$, is
given by
\begin{eqnarray}
\Pi(\nu + i\delta) & = & {ne^2 \over m}
\Biggl\{ -1 + \int_0^\infty d\omega \tanh({\beta\omega \over 2})
\Bigl(
h_1(\omega,\omega + \nu) -  h_2(\omega,\omega + \nu)
\Bigr)
\nonumber \\
& & \phantom{{ne^2 \over m}\Biggl\{ -1} + \int_{-\nu}^D
d\omega \tanh({\beta (\omega + \nu) \over 2})
\Bigl(
h_1^\ast (\omega,\omega + \nu) +  h_2(\omega,\omega + \nu)
\Bigr)
\Biggr\}
\label{kernel}
\end{eqnarray}
\noindent with
\begin{eqnarray}
h_1(\omega_1,\omega_2) & = & {1 - N(\omega_1) N(\omega_2) -
P(\omega_1) P(\omega_2) \over 2(\epsilon(\omega_1) + \epsilon(\omega_2)) }
\nonumber \\
h_2(\omega_1,\omega_2) & = & {1 + N^\ast(\omega_1) N(\omega_2) +
P^\ast(\omega_1) P(\omega_2) \over 2(\epsilon(\omega_2) -
\epsilon^\ast(\omega_1)) }
\nonumber \\
N(\omega) & = & { \wtilde(\omega +i\delta) \over \epsilon(\omega + i\delta) }
\nonumber \\
P(\omega) & = & { \phi(\omega +i\delta) \over \epsilon(\omega + i\delta) }
\nonumber \\
\epsilon(\omega) & = &
\sqrt{\wtilde^2(\omega +i\delta) -\phi^2(\omega +i\delta )}.
\label{definitions}
\end{eqnarray}
\noindent In eq. (\ref{kernel}) $D$ is a large cutoff of order the electronic 
bandwidth. \par

To summarize the previous set of equations: for a given model for the spectral
density, $\alpha^2F(\nu)$ and a choice of impurity scattering rate $1/\tau$,
we can compute the conductivity $\sigma(\nu)$ at any frequency and temperature,
including effects due to both elastic and inelastic scattering mechanisms, the
latter being determined by the choice of electron-phonon spectral density. 
As explained in the introduction, however, vertex corrections are
omitted.\par

In the analysis of experimental data it is possible to use several methods
to extract a single temperature dependent scattering rate. One such method
utilized the microwave conductivity in \YBCO \cite{bonn1,bonn2} and in 
Nb \cite{klein}, and adopted a two-fluid model description of the 
superconducting state.
It was assumed that the absorptive component of the conductivity (real part of
$\sigma$ at finite frequency, denoted by $\sigma_1(\nu)$) was due only to the
normal component of the fluid. Thus an expression of the form
\be
\sigma_1(\nu,T) = {ne^2 \over m} {m \over m^\ast(T)} 
\Biggl[ 1- {\lambda^2(0) \over \lambda^2(T)}
\Biggr] {\tau(T) \over 1 + (\nu\tau(T))^2}
\label{twofluid}
\ee
\noindent is assumed to hold approximately. In eq. (\ref{twofluid}) $\tau(T)$
has units of a scattering time and $\lambda(T)$ is the penetration depth
at temperature $T$. Here we will calculate $\sigma_1(\nu,T)$ for a model 
$\alpha^2F(\nu)$ and $1/\tau$ using the full expression (\ref{cond}). At
the same time we can calculate the penetration depth, either from a zero
frequency limit of eq. (\ref{cond}), or directly from the imaginary axis
\cite{nam,marsiglio90}:
\be
{\lambda^2(0) \over \lambda^2(T)} = \pi T \sum_{m=-\infty}^\infty
{\phi_m^2 \over (\omega^2_m Z^2(i\omega_m) + \phi_m^2)^{3/2} }.
\label{penet}
\ee
\noindent The idea is to examine the zero frequency limit of 
eq. (\ref{twofluid}), and thus define a scattering rate relative to the
rate at $T_c$ \cite{klein}:
\be
{\tau(T_c) \over \tau(T)} \equiv \Biggl(1 - {\lambda^2(0) \over \lambda^2(T)}
\Biggr)  {\sigma_N(T_c) \over \sigma_1(T)},
\label{klein_onet}
\ee
\noindent where it has been assumed that the mass enhancement factor in
eq. (\ref{twofluid}) does not change with temperature.\par

In Fig. 10 we show results for $\tau(T_c)/\tau(T)$ defined by 
eq. (\ref{klein_onet})
in the superconducting state (solid curve) with which we compare the
inverse lifetime for both the normal (short-dashed curve, 
eq. (\ref{gammak}) ),
and the superconducting (long-dashed curve, eq. (\ref{lifetime}) ) states.
The results are
based on a Debye model spectrum for $\alpha^2F(\nu)$ used in the previous 
section, with mass enhancement parameter $\lambda = 1$ and $T_c = 100$ K. (A
negative $\mu^\ast$ is required.) Note that the normal state scattering rate
(short-dashed curve) is only sensitive to the low frequency part of
$\alpha^2F(\nu)$ at low temperatures: a $\nu^2$ dependence in $\alpha^2F(\nu)$
implies a $T^3$ dependence in $1/\tau(T)$ (Again, vertex corrections
would alter this to the familiar $T^5$ law \cite{ashcroft}.)
The agreement between the inverse 
lifetime (long-dashed curve) and the scattering rate defined by the 
two-fluid model (solid curve) in the
superconducting state is remarkable. This indicates that the two-fluid
description makes sense \cite{berlinsky,bonn3}, and the quantities shown in 
Fig. 10 apply to the normal component of the superfluid. 
Fig. 10 illustrates the comparison in the clean limit, where the
two-fluid description is expected to be most accurate \cite{bonn3}. Before
investigating impurity dependence, we turn to a second possible procedure
for extracting a scattering rate from conductivity data \cite{romero,tanner},
which is simply a generalization of that used by Shulga {\it et al.} 
\cite{shulga} to the superconducting state. One simply fits the low frequency
absorptive part of the conductivity to a Drude form:
\be
\sigma_1(\nu,T) = {ne^2 \over m}{1 \over m^\ast/m} {\tau^\ast(T) \over
1 + (\nu\tau^\ast)^2}.
\label{romerofit}
\ee
\noindent As described in 
Refs. \cite{shulga,marsiglio95a}, 
it is possible to fit a Drude form to the low frequency 
part of the optical conductivity in the normal state. Such a fit is also
possible in the superconducting state \cite{romero}. Theoretically, the fit
is problematic in a BCS approach because there is a Hebel-Slichter
logarithmic singularity at low frequency {\it at all temperatures in the
superconducting state}. However, with the Eliashberg approach, the
Hebel-Slichter singularity is smeared, and one can fit a Drude form over
a limited range of frequency. Such a fit for a Debye spectrum is also included
in Fig. 10 (dotted curve). While the fit is in qualitative agreement
with the inverse lifetime, this method of characterizing the inverse lifetime
in the superconducting state is clearly not as accurate as the two-fluid
model. The fits themselves are shown in Fig. 11. It is
clear that the fits fail at sufficiently high frequency, as one would expect,
but that they characterize well the low frequency response in the 
superconducting state. \par

Fig. 10 clearly shows that there is considerable freedom and hence ambiguity
in extracting a temperature dependent scattering rate from conductivity data.
Nevertheless, in the clean limit it is evident that the two-fluid model
is a useful device for extracting the low energy quasiparticle inverse lifetime 
in the superconducting state.\par

With the addition of impurities the situation changes considerably. This is 
illustrated in Fig. 12, where the same calculations as in Fig. 10 are
shown, but with an additional impurity scattering, $1/\tau = 25$ meV
included. The use of formula (\ref{twofluid}), inspired by the 
two-fluid model,
gives a scattering rate (solid curve) that falls much less rapidly
around $T=T_c$ then does the inverse lifetime in the superconducting state
(long-dashed curve). The latter curve drops almost vertically as the 
temperature drops below $T_c$, as has already been discussed.
The result based on the Drude fit (dotted curve) shows a peak
which is reduced in size from that in the clean limit (Fig. 10). Indeed,
for increased impurity scattering, the peak just below $T_c$ disappears.
In both cases a rapid suppression is expected just below $T_c$: in the 
case of the inverse lifetime, this suppression is a consequence of Anderson's
theorem \cite{anderson}, as already discussed. In the case of the Drude fit,
it is easy to see that this is the case in the dirty limit. In the dirty limit
the conductivity  is almost flat as a function of frequency on the scale
of $1/\tau$, at $T_c$. Just below $T_c$, however, a gap in the spectrum
begins to develop, so that weight is shifted from roughly the gap region
to the delta function at the origin. Any low energy fit will then use
a Lorentzian width which senses this depression in the conductivity, which is
on an energy scale of the gap. This represents a significant suppression
from the normal state scattering rate (infinite in the dirty limit). Here
we are in an intermediate regime, with $1/\tau = 25$ meV (note: $\Delta(T=0)
= 20.2 $ meV). The corresponding fits for Fig. 12 are shown in Fig. 13.\par

%

Finally, we examine the low frequency conductivity vs. reduced temperature,
a quantity measured in the high $T_c$ oxides by microwave techniques. In
Fig. 14 we show the real part of the conductivity, $\sigma_1(\nu)$ vs.
reduced temperature, for several low frequencies. Note the relative 
insensitivity to frequency, a feature of strong coupling pointed out
in Ref. \cite{marsiglio91b}. Also note the lack of a coherence peak just
below $T_c$. Nonetheless, a broad peak exists at lower temperatures,
somewhat reminiscent of that observed in \YBCO \cite{nuss,bonn1,bonn3}.
This peak exists because of a competition between an increasing scattering time
(making $\sigma$ increase) and a decreasing normal component (making $\sigma$
decrease, particularly as $T \rightarrow 0$). We should note that this peak
is most prominent in the clean limit. As Fig. 13 indicates (see values at 
the intercept), the peak is absent for a sufficiently large impurity scattering
rate. \par

It is of interest to examine what dependence these results have on
the electron-phonon spectral function. As an extreme we utilize
a spectrum which is sharply peaked at some high frequency, and, in contrast
to the Debye model employed above,
coupling to low frequency modes is absent; such
a spectrum models a strong coupling to an optic mode. We choose a triangular
shape for convenience, starting at $\omega_0 = 34.8$ meV with a cutoff at
$\omega_E = 35.5$ meV. The coupling constant is chosen so that the
mass enhancement value is $\lambda = 1$, in agreement with that chosen for
the Debye spectrum. As was the case there, a negative $\mu^\ast$ is used to
give $T_c = 100$ K, and the zero temperature gap was found to be close
to the Debye value.\par

In Fig. 15 we plot the real part of the low frequency conductivity, 
$\sigma_1(\nu)$ vs. reduced temperature, now using the  triangular spectrum
for $\alpha^2F(\nu)$. In contrast to the result for the Debye model, the low
frequency conductivity is strongly frequency dependent at low temperatures.
In fact, for $\nu = 0$, the conductivity appears to diverge.  We believe that
at sufficiently low temperature, this curve will actually achieve a maximum
and approach zero at zero temperature, but we have been unable to obtain this
result numerically. Once again there is a competition between an increasing
scattering time and a decreasing normal density component as the 
temperature is lowered from $T_c$. Here, however, the increasing scattering
time appears to be overwhelming the decrease in normal fluid density. The
key difference with the Debye spectrum is that here the spectrum has a big gap, 
so that the lifetime is increasing exponentially with decreasing temperature,
already in the normal state. Recall that in the Debye case the increase
followed a power law behaviour with decreasing temperature. Since the decrease
in normal fluid density is always exponential, this term dominates
in the Debye case,
whereas in the case of the gapped spectrum the competition is subtle, and 
will depend strongly on the details of the spectrum (an electron-phonon
spectrum with a much smaller gap will yield a zero frequency conductivity
which approaches zero at zero temperature, for example).\par

The physics of this conductivity peak is different from what has been already
proposed for \YBCO. It has been suggested that the excitation spectrum
{\em becomes gapped due to the superconductivity}, i.e. a feedback
mechanism exists which creates a low frequency gap in the excitation 
spectrum as the superconducting
order parameter opens up below $T_c$. Such a scenario has been explored within
a marginal Fermi liquid scheme \cite{nuss,nicol1,nicol2}, and is also
consistent with thermal conductivity experiments \cite{yu}. Here the 
conductivity peak arises because the spectrum is already gapped, and
the scattering rate is sufficiently high at $T_c$ because $T_c$ itself
is fairly high. We should warn the reader that this mechanism requires
a ``fine tuning'' of the spectrum, i.e. a large gap is required.\par

Note that for any non-zero frequency the conductivity has a visible maximum, 
and quickly approaches zero at sufficiently low temperature. Nonetheless
this turn around occurs for yet another reason: as one
lowers the temperature the Drude-like peak at low frequencies gets
narrower while at the same time the magnitude of the zero temperature
intercept increases. For any given finite frequency, then, a temperature
is eventually reached below which this frequency is now on the tail
of the Drude-like peak. This means that
while the zero frequency conductivity
increases, that at any finite frequency will eventually decrease
as the width becomes smaller than the frequency.
So in this case the increase in scattering
lifetime still dominates the decrease in normal fluid density, but,
because we are fixed at a finite frequency, the conductivity
decreases. Note that frequencies of order 0.01 meV
($\approx 2.4$ GHz) are within the range of microwave frequencies that are
used in experiments.\par

To show this more explicitly  we illustrate in Fig. 16 the various 
scattering rates obtained with the gapped spectrum. These are to be compared
with those shown for the Debye model in Fig. 10. Clearly the quantitative
agreement between the scattering rate inspired by the two-fluid model 
and the
inverse lifetime that was seen in Fig. 10 with the Debye spectrum was
fortuitous. While a qualitative correspondence between these two entities
continues to exist with the gapped spectrum, they are no longer in 
quantitative agreement. We have verified that this is generically
true, by investigating other spectra, not shown here.

\par
\noindent {\bf IV.  SUMMARY}
\par
\medskip
We have investigated the quasiparticle lifetime in an Eliashberg s-wave
superconductor, generalizing earlier work \cite{kaplan} to include
impurity scattering as well. We find that the quasiparticle lifetime
becomes infinite at low temperatures, independent of the impurity
scattering rate, which we understand as simply a manifestation of
Anderson's theorem \cite{anderson}. Thus, on general grounds, within
a BCS framework, the scattering rate should collapse to zero
in the superconducting state.\par

We have also investigated two methods for extracting the scattering rate from
the low frequency conductivity. One relies on a two-fluid model
picture, and the other simply utilizes a low frequency Drude fit. Neither
should necessarily correspond very closely to the quasiparticle
inverse lifetime, and we find that in general they do not, quantitatively.
Qualitatively, however, the scattering rate defined by either procedure
gives the correct temperature dependence for the inverse lifetime. In the 
presence of impurities, the two-fluid prescription appears to be less accurate,
presumably because such a prescription takes into account only the lower
normal fluid density as the temperature is lowered, and not the fact
that impurity scattering is less effective in the superconducting state.
Thus we caution that interpreting a conductivity-derived
scattering rate as a quasiparticle inverse lifetime can lead to inaccuracies.
\par
\bigskip

\noindent {\bf ACKNOWLEDGEMENTS}
\par
F.M. wishes to acknowledge helpful discussions concerning the importance of 
quasiparticle lifetimes in detector applications with B. Sur. We thank
C. Kallin for some helpful suggestions.
This research was partially supported by the
Natural Sciences and Engineering Research Council (NSERC) of
Canada and the Canadian Institute for Advanced Research (CIAR).
\par

\vfil\eject
\bigskip
\bigskip
\noindent {\bf Figure Captions}
\par
\begin{itemize}
 \item Fig. 1. (a) Mass enhancement parameter, as defined by 
Eq. (\ref{lambdastar}) in the text, vs. reduced 
temperature $T/T_c$, for various
electron-phonon spectral functions. In all cases $T_c = 100$ K. The 4
spectra used are a Debye spectrum (solid line), linear spectrum (dotted line),
a spectrum proportional to $\sqrt{\nu}$ (dashed line), and a triangular
spectrum (dot-dashed line). In all cases the strength is such that 
$\lambda = 1$. In the first three cases a cutoff frequency equal to 30 meV
was used. In the last case, the spectrum starts at 34.8 meV and is cut off
at 35.5 meV.\par
(b) Inverse lifetime, $2\Gamma(T)$ (in meV) vs. reduced temperature for
the same spectra as in (a). Note the different low temperature behaviour, 
depending on the low frequency characteristics of the electron-phonon
spectrum used.\par

\item Fig. 2. Scattering rate, $2\Gamma(T,E)$ (in meV) vs. E (in meV) for
various temperatures in the superconducting state. The zero temperature
gap at the Fermi surface is indicated by the arrow. Note that for $T/T_c= 0.5$
the function plotted actually becomes negative (near 10 meV). The physical
pole occurs at an energy, E,  where $2\Gamma(T,E)$ is always positive, however.

\item Fig. 3. Scattering rate normalized to the zero temperature gap,
$2\Gamma(T,E)/\Delta$ vs. reduced temperature $T/T_c$, for various
quasiparticle energies, $\epsilon$. These are computed using the
perturbative expansion, Eq. (\ref{gammaimpbcs}). 
The impurity scattering rate is
$1/(\tau) \Delta = 1$. Note that on the Fermi surface (solid curve), the
scattering rate is zero immediately below $T_c$.

\item Fig. 4. (a) Real and (b) imaginary parts of the quasiparticle pole
vs. reduced temperature, $T/T_c$, with $1/(\tau) \Delta = 1$ and various
quasiparticle energies. These are computed non-perturbatively from
Eqs. (\ref{realpart},\ref{imagpart}). Note that  at the Fermi surface the 
quasiparticle energy has an abrupt onset at a temperature somewhat 
{\em below} $T_c$. The results in (b) are similar to those in Fig. 3, except
that at the Fermi surface, the decrease in scattering rate below $T_c$
is not as abrupt.

\item Fig. 5. The single particle spectral function, $A(k,\omega)$, vs.
normalized frequency, $\omega/\Delta$, with $1/(\tau) \Delta = 1$, for
various quasiparticle energies, $\epsilon/\Delta$. The Fermi surface
result (solid curve) has particle-hole symmetry. The result for 
$\epsilon/\Delta = 0.5$ would have a peak within the gap (between -1 and 1)
except that the coherence factors in Eq. (\ref{greennew}) give zero
residue for the gap region. As the quasiparticle energy increases, the spectral
function begins to resemble the normal state spectral function. Small
square-root singularities still exist, nonetheless at the particle and hole 
gap edges.

\item Fig. 6. The quasiparticle
scattering rate, $2\Gamma(T,E=0)$ vs. $T/T_c$ for various impurity
scattering rates, $1/\tau$, in both the superconducting and normal
(long-dashed curves) states. In the clean limit (solid curve) there is
an enhancement immediately below $T_c$. When impurity scattering is
present, this scattering is immediately suppressed in the superconducting
state, as shown by the dotted and dashed curves. Below a temperature of
about 0.8$T_c$ the scattering rate is independent of the amount of impurity
scattering present in the normal state. A Debye electron-phonon spectrum
was used with $\lambda = 1$ and cutoff frequency, $\omega_D = 30$ meV.

\item Fig. 7. The single particle spectral function, $A(k_F,\omega)$ at the
Fermi surface, vs.
frequency, in the clean limit, for various reduced temperatures. The spectral
function is considerably broadened near $T_c$, due to temperature alone. At
the two lowest temperatures shown, while there is no true gap in the excitation
spectrum, this plot makes it clear that, practically speaking, an effective
gap in the excitation spectrum is present. A Debye electron-phonon spectrum
was used with $\lambda = 1$ and cutoff frequency, $\omega_D = 30$ meV.

\item Fig. 8. The single particle spectral function, $A(k_F,\omega)$ at the
Fermi surface, vs.
frequency, for various impurity scattering rates, as indicated. Note the
broadening which occurs with increasing impurity scattering. However, spectral
weight remains absent in the ``gap region''.
Results are for a temperature $T/T_c = 0.5$, and with a Debye electron-phonon
spectrum with $\lambda = 1$ and cutoff frequency, $\omega_D = 30$ meV.

\item Fig. 9. The quasiparticle
scattering rate, $2\Gamma(T,E=\Delta(E))$, evaluated at the quasiparticle
energy, given on the Fermi surface by $E = \Delta(E)$,  vs. $T/T_c$ for 
various impurity scattering rates, $1/\tau$, in both the superconducting 
and normal (long-dashed curves) states. These results are in quantitative
agreement with those in Fig. 6, since the energy scale, $\Delta$, is
still small compared to other (phonon) energy scales in the problem.
In particular, below 
about 0.8$T_c$,  the scattering rate is independent of the amount of impurity
scattering present in the normal state. A Debye electron-phonon spectrum
was used with $\lambda = 1$ and cutoff frequency, $\omega_D = 30$ meV.

\item Fig. 10. Various normalized scattering rates vs. reduced temperature.
The normal and superconducting scattering rates come from the quasiparticle
inverse lifetime, given by eq. (\ref{lifetime}), at zero energy. The two-fluid
result comes from eq. (\ref{klein_onet}) while the Drude fit is obtained
by fitting eq. (\ref{romerofit}) to the low frequency conductivity in the
superconducting state. Note the agreement of the scattering rate as
extracted from the two-fluid analysis with the inverse lifetime in the
superconducting state. A Debye electron-phonon spectrum
was used with $\lambda = 1$ and cutoff frequency, $\omega_D = 30$ meV.

\item Fig. 11. The low frequency conductivity in the superconducting
state (solid curves) along with their fits based on  eq. (\ref{romerofit})
(dashed curves). These fits were used in Fig. 10 (dotted curves).

\item Fig. 12. Same as Fig. 10, except now with an impurity scattering rate
of 25 meV. Note that the two-fluid analysis agrees poorly with the 
quasiparticle inverse lifetime.

\item Fig. 13. The low frequency conductivity fits used in Fig. 12.

\item Fig. 14. The very low frequency conductivity as a function of
reduced temperature, in the clean limit.  A Debye electron-phonon spectrum
was used with $\lambda = 1$ and cutoff frequency, $\omega_D = 30$ meV.
Note that the results are relatively  insensitive to frequency (the
$\nu = 0.01$ meV result, given by the dotted curve, is essentially hidden
by the zero frequency result). A microwave
experiment yields essentially  zero frequency results.

\item Fig. 15. Same as for Fig. 14, but now with the triangular spectrum
as described in the text. Note that the zero frequency conductivity appears
to diverge as $T \rightarrow 0$. A microwave experiment will yield a large
low temperature peak as a function of reduced temperature, whose magnitude
will depend strongly on the frequency. These results are for the clean limit.
The peak is also reduced as impurity scattering is added (not shown).

\item Fig. 16. Comparison of scattering rates (as in Fig. 10) calculated 
for the triangular spectrum, in the clean limit. Note that the normal
state result approaches $T=0$ exponentially due to the gap in the 
$\alpha^2F(\nu)$ spectrum (in Fig. 10 the corresponding curve approached
zero with a power law behaviour). While the results in the superconducting
state are qualitatively similar, they no longer agree quantitatively with
one another, as in Fig. 10.

\end{itemize}

\begin{thebibliography} {999}
\bibitem{comm}
It is true that we can attempt a representation through Pad\'e approximations,
as has often been done in the past to analytically continue functions from
the imaginary axis to the real axis. However, these are notorious for
producing spurious poles \cite{msc}.
\end{thebibliography}
\end{document}